\documentclass[12pt]{article}
\usepackage{amsmath,amssymb,eepic,float,epsfig}
\begin{document}

\title{Relaxation of twisted vortices\\ in the Faddeev-Skyrme model} 
\author{Jarmo Hietarinta$^1$, Juha J\"aykk\"a$^1$, and Petri
  Salo$^2$\\
$^1$Department of Physics, University of Turku,\\
 FIN-20014 Turku, Finland\\
$^2$Laboratory of Physics,
 Helsinki University of Technology,\\ P.O. Box
1100, FIN-02015 HUT, Espoo, Finland} 
\date{\today}\maketitle

\begin{abstract}
  We study vortex knotting in the Faddeev-Skyrme model.  Starting with a
  straight vortex line twisted around its axis we follow its evolution under
  dissipative energy minimization dynamics. With low twist per unit length the
  vortex forms a helical coil, but with higher twist numbers the vortex
  becomes knotted or a ring is formed around the vortex.
\end{abstract}



Vortices are familiar objects in the physical world, from smoke rings
to hurricanes and maelstroms. There is always an associated vector
field, which  in the above-mentioned examples is the velocity field of the
underlying material, and which has a special value that determines the core
of the vortex (e.g., the eye of a hurricane).  But by its nature, 
velocity-vector fields can deform continuously to zero and therefore the
corresponding vortex can disappear.

Recently there have been experimental observations of vortices in more exotic
circumstances in which the relevant vector field is not associated to velocity
but rather to spin or other such property.  Then it is possible to have
vortices that are both {\em nonsingular} and {\em conserved}, and the
conservation follows from topological reasons. A typical example of such a
case is a vortex in a material that is described by a unit-vector field ${\bf
  n} = (n_1,n_2,n_3)$, $|{\bf n}|=1$.  There are many different topological
properties that could come to play, but here we are only interested in vector
fields that are characterized by the Hopf charge.  Then we have a nonsingular
unit vector field for which one can {\em locally} define a topological charge
density whose integrated value is conserved. Thus, in this case, we assume
that faraway the vector field is parallel, which is in contrast to
topological charges determined from nontrivial asymptotic properties as is the
case with monopoles.  Physically such vortex configurations can be obtained
for example in the ``continuous unlocked vortex'' (CUV) structure of rotating
superfluid ${}^3$He-A \cite{rota}: If the vortex is along the $z$-direction,
then far away in the $(x,y)$-plane the $\hat{\bf l}$-spin points to the
$x$-direction, at the vortex core to the opposite direction, and in between
the 3D unit vectors interpolate these values  continuously.  In general,
topological vortex structures have been detected in liquid helium in the
superfluid state \cite{rota, CosLab1, CosLab2}, two dimensional electron gas
systems like the quantum Hall ferromagnets \cite{hall}, nematic liquid
crystals \cite{nema}, Bose-Einstein condensates \cite{bec} and superconductors
\cite{supercond}. They probably exist also in cosmic strings \cite{Cosmic},
the solar corona \cite{solar} and in SU(2) Yang-Mills model \cite{su2ym}.  In
many of these examples the unit vector field mentioned above is not manifest,
but it can be seen when the standard physical variables are suitably
parameterized \cite{extract}.

Vortices in 3D are often constructed simply by stacking identical
$(x,y)$-layers of 2D vortex structures. In the CUV model mentioned above, the
asymptotic behavior and core location are invariant under a global rotation of
the vector field around the $x$-direction, and the model itself is often
invariant under such a gauge rotation as well. In that case, one can construct
{\em twisted vortices}: the vortex core (e.g., the $z$-axis) is straight, the
asymptotic direction of the vortex field (in the $(x,y)$-plane) is constant
(e.g., along the $x$-axis), but the vector field on different $z$-layers is
obtained from the initial 2D vortex by a $z$-dependent gauge rotation around
the the asymptotic direction of the vector field.  One hopes that such twisted
vortices can also eventually be made in the laboratory, and that they can be
seen to play a role in some of the physical processes mentioned above.

Once twisted vortices have been created one may ask about their stability, and,
if they are unstable, how they tend to deform.  Rubber-band experiments
illustrate that the cost balance between twisting and stretching is sometimes
such that it is advantageous to convert twisting into stretching.  One may
therefore expect that, when the vortex twist per unit length is high enough
and the system is allowed to evolve into its nearest energy minimum, the
resulting state may involve a bent or even knotted vortex core with less
twisting along the core.

In this Letter we present the results of a study on twisted vortices
in the Faddeev-Skyrme model.  Previous numerical simulations of this
model have concentrated on stable symmetric rings with the Hopf charge
of 1 and 2~\cite{FN,GH}, on bent rings (un-knots) \cite{BS}, and on
linked un-knots and rings with multipole core \cite{HS99, HS00, HJS03,
  Vid}.  In these studies the model has been shown to support various
kinds of knotted vortex solitons with nonzero Hopf charge.  Here we
study the relaxation process and minimum-energy solutions of twisted
vortices.  The results will be helpful in verifying the existence of
twisted vortices and hopefully also in constructing knotted structures
in various media.

The Faddeev-Skyrme model is defined by Faddeev's Lagrangian~\cite{Fa} 
\begin{equation}
L=\tfrac12\int\left[(\partial_\mu {\bf n})^2 + g
F_{\mu\nu}^2\right] d^3x,
\quad F_{\mu\nu}= \epsilon_{abc}
n^a\partial_\mu n^b \partial_\nu n^c,\quad {\bf n}^2=1.
\label{FL}
\end{equation}
If we assume that the unit vector field ${\bf n}$ has a fixed value
${\bf n}_\infty$ in all asymptotic spatial directions (which in the
vortex case means that it closes eventually) we can compactify 
${\mathbb R}^3$ (i.e., contract the spatial infinity to a point) 
and then the conserved topological charge is given
by the Hopf charge $Q$. It characterizes the homotopy of the mapping
${\bf n}: {\mathbb R}^3 \cup \{\infty\}\approx S^3\to S^2$, which is given by
$\pi_3(S^2)={\mathbb Z}$, and therefore these maps are classified by an
integer, which we call $Q$.

The basic definition of Hopf charge is designed for finite-size
objects, and it must be modified for vortices extending to infinity.
We consider here only periodic vortices and adapt the definition
straightforwardly to that case.  The Hopf charge can be computed from
its local differential-geometric definition
\begin{equation}
Q=\frac1{16\pi^2}\int \epsilon^{ijk} A_i F_{jk}\, d^3x\label{Q},
\end{equation}
where $A_i$ is defined via $F_{ij}=\partial _iA_j-\partial _jA_i$ and 
which can be implemented numerically and computed for one $z$-period.
An equivalent definition is based on the linking number of the
preimages of two different points on the target space $S^2$ of the
map ${\bf n}$ \cite{Bott} (see also \cite{HJS03}). This second
definition is also directly applicable to our periodic case.

In this work the initial configurations were constructed from 2D
vortex configurations, having the value ${\bf n}_\infty=(0,0,1)$ far
away from the $z$-axis and ${\bf n}_{core}=(0,0,-1)$ at the $z$-axis.
These 2D layers were stacked with a $z$-dependent twist, in the cylindrical
coordinates ($x=\rho\cos\theta, y=\rho\sin\theta$) we have
\begin{equation}\label{eq:vorteksinkaava}
  {\bf n}(x,y,z; n, m) = 
  \begin{pmatrix}
\sqrt{1-f(\rho)^2}\,\cos(m\theta+2\pi n z/L)\\
\sqrt{1-f(\rho)^2}\,\sin(m\theta+2\pi n z/L)\\
f(\rho)
  \end{pmatrix}.
\end{equation}
Here $L$ is the box length in the $z$-direction, and $f(\rho)$ is some profile
function with $f(0)=-1,\,f(\infty)=+1$, for example
$f(\rho)=1-2a/(a-1+e^{b\rho ^2})$. The values of $a$ and $b$ were selected to
give the initial configuration as low energy as possible.  In order to avoid
singular symmetric situations we introduced a small bend to the vortex
core-line.

The minimum energy configurations were obtained numerically using the
steepest-descent method, which was improved by taking into account
also the gradients of the previous step. The system was discretized on
a rectangular lattice, with grid sizes from $240^2 \cdot 120$ to
$480^2\cdot 240$. The corresponding box size is $5.0$ units in the
shortest dimension (for details of our computations in cubic lattices,
see \cite{HS99}).  At the $z$-boundaries we used periodic boundary
conditions, so that we had an integer amount of twists within the box.
At the $(x,y)$-boundaries the vector field was fixed to its vacuum value
${\bf n}_\infty$. The existence of such a rigid boundary could in
principle affect the evolution of the system, and therefore we used as
large a box as possible. The sufficiency of the discretization density
was followed by checking that the angle between two nearest-neighbor
vectors never exceeded $30^\circ$: if such a situation was approaching
the system was put into a denser grid.  
Computations were terminated whenever the
system temperature was low enough.\footnote{The present computations 
were done on CSC~-~Scientific
Computing's IBM eServer Cluster 1600 (16 pSeries 690 nodes of 32
POWER4 CPUs at 1.1 GHz each) and IBM SP (32 WinterHawk II nodes of 4
POEWR3-II CPUs at 375 MHz each) parallel supercomputer clusters. Each
round of iteration took about 15 milliseconds for the $240^2\cdot 120$
grid on the (faster) pSeries 690 computer and 360 milliseconds for the
$480^2\cdot 240$ grid on the (slower) SP system. The total number of
iterations was typically somewhat less than 200.000.}

We studied cases where the box contained 5 to 8 full twists along a single
vortex, lower twist-per-unit-length initial states tend to form helical
configurations without any interesting knotting.  (For further discussions on
the straight vortices of the Faddeev-Skyrme model, see \cite{sv, lubcke}.) In
Figures 1-4 we present snapshots of the deformation process for the cases
$Q=5\ldots8$ that were studied in detail.  In these Figures we have used the
``isosurface view'' and plotted a small isosurface around the ${\bf
  n}_{core}=(0,0,-1)$ value, the coloring describes vector orientation on this
surface, as described in \cite{HS99}.

In all cases the vortex string tends to lengthen and bend in space in
order to decrease twist along the core, in the leftmost parts of
Figures 1-4 this has already progressed to some extent.  If there is
enough twist per unit length some sections of the coil will eventually
touch and pass through each other thereby changing the topology of the
core line, this is illustrated, e.g., in Figure 2 (b) and (c). In
Figure 3 this process takes place twice and the result is a knot in
the vortex line.  In Figure 4 the process itself is more intricate,
but the result is also a knotted vortex core. The energies of the
final states are as follows (subscript stands for the Hopf charge)
$E_5= 506.44,\, E_6= 584.02,\, E_7=657.57 ,\, E_8=732.97$.  In this
range of $Q$ values the data can be fitted to the VK-bound \cite{VK}
$E_Q=152.91 |Q|^{3/4}$ (within 1\%), but a linear fit to $E_Q=130.72+
75.312 |Q|$ is even better (within 0.25\%).

During minimization the whole vector field changes slowly and the
behavior of the core line (which is the preimage of the south-pole of
the target space of $\bf n$) only tells part of the story.  Note,
e.g., that the deformation process is continuous and preserves the
Hopf charge, but nevertheless the knottedness of the core line
changes.  Thus in addition to the core line we should also follow the
twisting around it, indeed the proper mathematical description is
obtained using ``framed links''. In Figures 1-4 the twisting is
described by the colors, but another point of view is obtained if we
plot a {\em ribbon} whose center is at the core line \cite{HJS03},
i.e., the preimage of a short line-segment passing through the south
pole of the target space (actually the preimage of any line segment
carries the same information).  In Figure 5 we have plotted the data
used in Figure 2 using this ``ribbon view'' (for technical reasons the
ribbon is represented by 5 nearby narrow tubes).  We can see that in
this particular case a one-component ribbon will cross itself and
break into a twisted ring surrounding a less twisted ribbon. The
details of the crossing process is given in Figure 6 (see also Figures
7-10 in \cite{HJS03}).

The essential feature of the ribbon process is that it converts between a
twist in the ribbon and a crossing.  The process is such that the Hopf charge
is conserved, this is easily seen when we use the definition based on the
crossing and twisting of the ribbon, or equivalently on linking numbers of two
preimages (see \cite{HJS03}, Sec.~6-7). For this purpose the curves have to be
directed (for multicomponent curves the directions have to be consistent so
that when the target points are moved and number of components change we do not
have to reverse any directions). The curves of the different preimages cross
on the projected figure, and at each crossing we assign a $+1$ or $-1$
depending whether the (directed) over-crossing curve has to be rotated
counterclockwise or clockwise, respectively, to be parallel with the
under-crossing curve. The linking number is one half of the sum of these
crossing numbers. In Figure 7 we have used this ``preimage view'' and plotted
three pairs of curves corresponding to various preimage pairs of the final
state with $Q=5$. In each case the linking number is found to be 5.

One interesting observation is that the vortex core of the final state in
Figure 1 is not a simple curve, but seems to have a crossing.  A similar
situation is seen in the linked un-knots $1+2+2$ and $1+3+2$ in Figure 5 of
\cite{HS00}.  One might hope that it is just due to the finite thickness of
the plotted core-tube and it would disappear if we plotted a sufficiently
narrow tube, but this is not necessarily the case.  What really takes place is
again clarified in the ribbon view in Figure 8: In the presented final state
the core-line seems to be in the middle of a ribbon going through the
splitting process described in Figure 6.  Indeed, such crossed pre-image
curves are always present as intermediate states between preimages with a
different number of components: if one connects these preimages in the target
space by a curve then at least one point on this curve will interpolate
between preimages with different number of components, which is only possible
if its own preimage is a crossing curve.  (Furthermore, the collection of
points whose preimages are crossed curves will themselves form a continuous
curve on the target space, this will be discussed in a future publication.)

In conclusion, we have studied twisted vortices in the Faddeev-Skyrme model
and found how they deform under energy minimization. Twisting the vector field
more than four times around the vortex core (with the chosen dimensions) makes
the vortices unstable and the final configurations will be knotted or
otherwise deformed.  We have analyzed the deformation processes with different
views of the data.  The isosurface view, which emphasizes the vortex core
position, has been augmented with the ribbon view, which adds further
information on the details of the vector field during the deformation process.
In the preimage view one can see how the Hopf charge, which is conserved under
deformation, can be calculated from the linking of the different preimages
(even when they have different number of components).  Our numerical
calculations show that in this model the deformation is a complicated but
still well-defined process from a given initial twisted vortex to the final
configuration with the same Hopf charge. One hopes that the corresponding
structures can also be seen experimentally in some physical systems.

\section*{Acknowledgments}
We acknowledge the generous computer resources of CSC -- Scientific
Computing Ltd., Espoo, Finland.  This work has been supported in part
by the Academy of Finland, through project 47188 (J.H. and J.J.) 
and through its Center of Excellence program (P.S.). One of us (J.J.) 
would also like to thank COSLAB Programme for support. We would also like to
thank E.~Haviola for help with graphics.

\pagebreak

\section*{Figure captions}
\ 

Figure 1: Different stages in the deformation process from a straight twisted
vortex of charge $Q=5$. In the first snapshot the
vortex has already twisted considerably, in subsequent pictures different
parts of the core line touch and deform, forming a ring attached to the vortex
core.

Figure 2: As in Figure 1, but for $Q=6$. Now the final state is a vortex of
charge $Q=2$ surrounded with a ring with $Q=2$, together having $Q=6$.

Figure 3: As in Figure 1, but for $Q=7$. Now the final state is a knotted
vortex, it has been reached by two crossing processes. 

Figure 4: As in Figure 1, but for $Q=8$. The deformation process is still more
complicated, the result is the same knot as in Figure 3 but the vortex core
carries more twist.

Figure 5: The ribbon view of the vector fields in Figures 2 (b) and
(c), with an intermediate state. In the first picture ribbon edges at
two different places have already touched and the purple edge has gone
through the crossing-splitting deformation. In the middle picture the
deformation has gone through half of the ribbon, the center part
forming a cross. In the last picture the red edge is finally about to
form a crossing thereby finishing the process in which a ribbon ring
is formed around the vortex ribbon.

Figure 6:  Details of the ribbon crossing process. We have chosen the
ribbons to look identical at the boundaries of the figure and since
the ribbons interact through edges of same color one of them has been
partially twisted $180^\circ$. Whenever lines cross they rearrange and the
process goes through the whole ribbon. As a result we have two
non-crossing ribbons, one of which has developed a $360^\circ$ twist
while the other one can be straightened.

Figure 7: Preimages of three arbitrary pairs of points for the final state of
charge $Q=5$ vortex.  The Hopf charge can be computed from the linking numbers
of two different preimages, when curves are consistently directed, and the
result is independent of the number of components.

Figure 8: Final state of the $Q=5$ case in the ribbon view. The vortex
core  (colored grey) is in the middle of a splitting ribbon.

\end{document}